\documentstyle[preprint,aps]{revtex}
\begin{document}
\draft
\preprint{\vbox
{\hbox{SNUTP 94-08} \hbox{hep-ph/9404214} \hbox{March 1994} }}
\title{Instanton Contribution to \\
        $B \rightarrow X_u e \bar \nu $ Decay}
\author{Junegone Chay\footnote{E-mail address:
\tt chay@kupt.korea.ac.kr}
and Soo-Jong Rey\footnote{E-mail address:
\tt sjrey@phyb.snu.ac.kr}}
\address{Physics Department, Korea University, Seoul 136-701,
Korea${}^*$ \\
  Physics Department and Center for Theoretical Physics \\
  Seoul National University, Seoul 151-752, Korea${}^\dagger$ }

\date{\today}
\maketitle

\begin{abstract}
We study instanton effects on inclusive semileptonic
$b \rightarrow u$ decay using the heavy quark effective
field theory and the operator product expansion.
The effect contributes not only to the hadronic matrix element
but also to
the coefficient functions in the operator product expansion.
We find that the
coefficient function is singular near the boundaries of the phase
space.
In order to use perturbative QCD reliably, it is necessary to
introduce smearing near the boundary.
However the instanton contribution to $b \rightarrow c$ decay seems
negligibly small.
\end{abstract}
\narrowtext
\section {Introduction}
The Cabibbo-Kobayashi-Maskawa matrix element $V_{ub}$ may
be extracted from the shape of electron energy spectrum in
inclusive semileptonic $B$ decay.
In order to avoid large $b \rightarrow c$ decay background,
we have to
look at the endpoint region of ${\cal O}(300 {\rm MeV})$.
However this is precisely the regime in which theoretical
prediction is
least understood due to various low-energy QCD effects. Therefore
experimental data have been compared mainly with phenomenological
models such as ACCMM~\cite{accmm} or ISGW~\cite{isgw}.

Recently there has been important theoretical progress in
inclusive semileptonic $B$ decay. The first model-independent QCD
approach
has been formulated using the heavy quark effective field
theoy(HQEFT) and the operator product expansion(OPE).
Chay, Georgi and Grinstein~\cite{chayetal} have shown that the
weak decay of
$B$ mesons can be described by a systematic expansion in
inverse powers of
$m_b$, in which the leading contribution reproduces the
parton-model result (See also Ref.~\cite{bigietal}.).
Using the OPE, Chay et.al. have shown that next-order
corrections are of order $(\Lambda /m_b)^2$,
where $\Lambda$ is the QCD scale, hence expected to be small.
Mannel~\cite{mannel}, Manohar and Wise~\cite{manoharwise}
have analyzed these corrections and concluded that these
corrections are singular at the boundaries
of the Dalitz plot, which necessitates some smearing prescription
before comparing with experimental data.

There are also radiative QCD corrections to the decay
rate~\cite{corboetal}.
In the leading log approximation, double logarithms are
summed to provide the Sudakov factor.
In $b \rightarrow u$ decay, this changes significantly
the shape of the electron spectrum near the endpoint.
Falk, Jenkins, Manohar and Wise~\cite{falketal}
have considered the validity of this resummation near the
endpoint of electron energy.
By studying the pattern of the summation of subleading logarithms,
they have reached a conclusion that a smearing is necessary for
the validity of the leading log approximation.
Bigi et.al.~\cite{bigiope} have reached essentially
the same conclusion but with a careful study of running
$\alpha_s (k^2)$.

The idea of using the OPE in the approach of Ref.~\cite{chayetal}
is that it factorizes the short-distance and the long-distance
physics in terms of the product of the Wilson coefficient
functions and the matrix elements of local operators.
So far the main focus of recent discussions has been on the
systematic analysis  of nonperturbative hadronic matrix elements.
However the coefficient functions may be affected by
nonperturbative effects as well.
One of the examples is an instanton contribution.

In this paper we investigate the effect of instantons on the
coefficient functions in $B$ decay.
Due to asymptotic freedom $\alpha_s (Q^2) \sim 1 / \ln Q^2$
as $Q^2 \rightarrow \infty$, we expect that the strength of
a single instanton is of order $\exp(-2 \pi / \alpha_s (Q^2))$.
At large momenta this is negligible compared to perturbative
QCD corrections.
However the characteristic scale for $\alpha_s$ is not $m_b$
but the momentum transfer to final hadrons.
Therefore near the endpoint of electron energy spectrum
where $\alpha_s$ becomes large, it is likely that instanton
effects on the decay rate may grow sizably.
For large momentum transfer $Q^2$ to final hadrons,
we expect that small instantons of size $\rho
\mathrel{\rlap{\lower3pt\hbox{\hskip0pt$\sim$}}\raise1pt\hbox{$<$}}
1/|Q|$ are most relevant.
We study their effect by calculating one instanton
(+ anti-instanton) corrections to the inclusive semileptonic
$b$ decay rate.
We find that the effect is negligible in most of the
the phase space except near the boundaries of the Dalitz plot.
At the boundaries we find that the instanton contribution
becomes singular.
Therefore the decay rate has to be smeared in such a way that
the instanton contribution remains small compared to the
parton-model result.

In this estimate we use the dilute gas approximation since the
main contribution comes from small instantons.
Instantons contribute not only to the hadronic matrix elements
but also to the coefficient functions.
We emphasize that we have neither attempted to calculate QCD
radiative corrections in the realistic instanton background
nor other nonperturbative effects such as multi-instantons or
renormalons.
It is possible that any of these give bigger contributions
than a single instanton contribution.
Even in this case our estimate may still characterize a typical
size of nonperturbative effects in the coefficient functions.
After all the dilute instantons are the only known
calculable nonperturbative effect in QCD.

This paper is organized as follows. In Section 2 we discuss
the kinematics relevant to the inclusive semileptonic $B$ decay.
In Section 3 we recapitulate the OPE and summarize the general
structure of local operators and their coefficient functions.
In Section 4 we compute the instanton contribution to the
coefficient functions in the forward Compton scattering amplitude.
In Section 5 we discuss numerical estimate of the results of
Section 4.
We conclude in Section 6  and also comment on the nonperturbative
effects in other cases.

\section{Kinematics}
The semileptonic $b \rightarrow u $ decay is described by the weak
Hamiltonian
\begin{eqnarray}
\displaystyle
H_W & = & \frac{G_F}{\sqrt{2}} V_{ub} {\overline u} \gamma^{\mu}
(1+\gamma_5 )b \cdot {\overline e}\gamma_{\mu} (1+\gamma_5) \nu_e
\nonumber \\
& = & \frac{G_F}{\sqrt{2}} V_{ub} \, J_u^{\mu} J_{e \mu}^{\dagger},
\label{ham}
\end{eqnarray}
where $J^\mu_u = \bar u (1 + \gamma_5) b$ and
$J^\mu_e = \bar \nu_e \gamma^\mu (1 + \gamma_5) e$.
For $b \rightarrow c$ decay the Hamiltonian is obtained from
Eq.~(\ref{ham}) by replacing $u$ and $V_{ub}$ by $c$ and
$V_{cb}$ respectively.
The inclusive decay rate is related to the hadronic tensor
defined by
\begin{equation}
W^{\mu \nu} = (2\pi)^3 \sum_X \delta^4 (p_B -q - p_X) \langle B |
J_u^{\mu \dagger} |X\rangle \langle X | J_u^{\nu} |B\rangle,
\label{wmunu}
\end{equation}
where $X$ denotes all possible hadronic final states.
Let $k^\mu_e$ be the electron momentum, $k^\mu_{\bar \nu}$ be the
antineutrino momentum, and $p^\mu_B = M_B v^\mu$ be the momentum
of the $B$ meson.
Then the hadronic tensor $W_{\mu \nu}$ depends on the timelike
vector $q^{\mu} = k^{\mu}_e + k^{\mu}_{\bar \nu}$.
In this paper we do not distinguish the $B$ meson mass $M_B$ and
the $b$ quark mass $m_b$.
This difference may be incorporated into the kinematics as
discussed in Refs.~\cite{bigietal,manoharwise}.

The differential decay rate of the $B$ meson is given by
\begin{equation}
d\Gamma = \frac{G_F^2}{4 m_b} |V_{ub}|^2 W^{\mu \nu} L_{\mu \nu}
d({\rm {PS} }),
\label{differential}
\end{equation}
where $L_{\mu\nu}$ is the leptonic tensor given by
\begin{equation}
L^{\mu \nu} = 8 (k^{\mu}_e k^{\nu}_{\bar \nu} +
k^{\nu}_e k^{\mu}_{\bar \nu} -g^{\mu \nu} k_e \cdot k_{\bar \nu}
-i{\epsilon^{\mu\nu}}_{\alpha\beta} k^{\alpha}_e
k^{\beta}_{\bar \nu} ).
\label{lmunu}
\end{equation}
We can express the phase space integral in terms of the three
independent
kinematic variables $ y = 2 E_e / m_b$, $\hat q^2 = q^2/m_b^2$
and $v \cdot \hat q = v \cdot q / m_b$ as
\begin{equation}
d ( {\rm PS})
= {d^3 k \over (2 \pi)^3 2 E_e} {d^3 k' \over (2 \pi)^3 2 E_\nu}
= {m^4_b \over 2^7 \pi^4} \,\, d v \!\cdot\!\hat q \,\,dy\,\,
d \hat q^2.
\label{psint}
\end{equation}

In the complex $v \cdot \hat q$ plane with $\hat q^2$ fixed,
$W^{\mu \nu}$ is related to the discontinuity of the forward
Compton scattering amplitude $T^{\mu \nu}$ across a physical cut as
\begin{equation}
W^{\mu\nu} = 2 \, {\rm Im} \, T^{\mu\nu},
\end{equation}
where
\begin{equation}
T^{\mu\nu} (q, v) = -i \! \int d^4x \, e^{i q \cdot x} \,
\langle B | T\{J_u^{\mu}(x)^{\dagger} J_u^{\nu}(0)\} |B \rangle.
\label{tmunu}
\end{equation}

To evaluate the decay rate it is necessary to examine the
analytic structure
of $T^{\mu \nu}(q, v)$ ~\cite{chayetal,manoharwise}.
In the complex $v\cdot\hat q$ plane, a ``physical cut'' relevant
to the decay is located on the real axis between
$\sqrt{\hat q^2} \le v\cdot \hat q \le (1 + \hat q^2 -\rho )/2$
where $\rho \equiv m_u^2 / m_b^2$ is the mass ratio-squared of
the $u$ quark to the $b$ quark.
Discontinuity of $T^{\mu \nu}$ across this physical cut
yields $W^{\mu\nu}$ for semileptonic $B$ decay. There are also
other cuts for $v\cdot \hat q \le -\sqrt{\hat q^2}$ and
$v \cdot \hat q \ge \frac{1}{2} ((2 + \sqrt \rho)^2 - \hat q^2 -1)$
but they correspond to other physical processes.
In Fig.1 we show the analytic structure of $T^{\mu \nu}$ in
$v \cdot\hat q$ plane.
As $v\cdot \hat q$ approaches $(1 + \hat q^2- \rho^2)/2$ the
final hadron reaches a resonance region.
There we expect significant nonperturbative QCD effects.
Chay et.al.~\cite{chayetal} have observed that one can calculate
the decay rate even in this limit perturbatively by choosing
the integration contour away from the resonance region.
This should provide a good approach to $b \rightarrow c$
inclusive decay.
Since $m_c^2 \gg \Lambda^2$, the other cut on the right-hand
side in Fig.1 always stays away from the physical cut.

However this is no longer true in $b \rightarrow u$ decay.
In this case we cannot choose a contour away from the resonance
region without enclosing some of the unphysical cut.
When $\hat q^2 \rightarrow 1$ and $\rho \rightarrow 0$,
the physical cut and the right-hand side cut in Fig.1 pinch
together. This is in fact the origin of large logarithms of
lepton energy found in Ref.~\cite{corboetal}.
Since the charm threshold lies at $y = 1 - \rho \sim 0.9$,
we have to deal with this problem.
For the moment we consider $q^2 \ll m_b^2 $ but
$(m_b v - q)^2 \gg \Lambda^2$ in order to calculate the decay
distribution reliably using perturbative QCD.
In the end we are interested in how far the result may be
extended to the endpoint region.

The total decay rate is given by
\begin{equation}
\Gamma = {G_F^2 m_b^5 \over 2^9 \pi^4} |V_{ub}|^2 \int_0^1 d y \,
\int_0^y \! d \hat q^2
\int_{{y \over 2} + {\hat q^2 \over 2 y}}^{{1 + \hat q^2 \over 2}}
\!\! d \hat v \cdot q \, {W^{\mu \nu} L_{\mu \nu} \over m_b^2}.
 \label {total1}
\end{equation}
Note that the interval of $v \cdot \hat q $ integration in
Eq.~(\ref{total1}) does not coincide with the physical cut.
{}From the kinematics the minimum of
$v \cdot \hat q$ is at point $P$ in Fig.1 where
$v \cdot \hat q = y / 2 + \hat q^2 / 2 y$.
In Eq.~(\ref{total1}) $W^{\mu \nu} L_{\mu \nu}$ is evaluated from
the discontinuity of $T^{\mu \nu} L_{\mu \nu}$ along the contour
$C^{\prime}$ in Fig.1.
Using Cauchy's theorem this is related to the contour integral
along the contour $C$ in the complex $v \cdot \hat q$ plane.
The integral along $C$ is reliably evaluated using perturbative
QCD as long as $(m_b v - q)^2 \gg \Lambda^2$.
Then the double differential decay rate can be expressed as
\begin{equation}
{d^2 \Gamma \over d \hat q^2 d y} = - {G_F^2 m_b^5 \over 2^8 \pi^4}
|V_{ub}|^2
\int_C \! d v \cdot \hat q \,\, {T^{\mu \nu} L_{\mu \nu} \over m_b^2}
\label{doublerate}
\end{equation}
The contour $C$ is shown in Fig.1. Care must be taken in
choosing the contour $C$ in Eq.~(\ref{doublerate}).
For example if $T^{\mu \nu}$ contains poles only,
any contour enclosing the poles gives the right
answer~\cite{chayetal,manoharwise}.
On the other hand if $T^{\mu \nu}$ has cuts, the contour has to be
chosen in such a way to cover the correct $v \cdot \hat q$
integration interval determined by kinematics.
We therefore have chosen the contour $C$ in
Eq.~(\ref{doublerate}) as
a circle of radius $z  = (y - \hat q^2) ( 1 - y) / y$ centered at
$v \cdot \hat q = (1 + \hat q^2 )/2$.
The contribution of the contour integral near the point $P$,
$ v \cdot \hat q = y/2 + \hat q^2 / 2y$
is negligibly small as long as  $z \gg \Lambda / m_b$.

Near the boundaries of the Dalitz plot $\hat q^2 \rightarrow y$ or
$y \rightarrow 1$, $z$ shrinks to zero and we are in the
resonance region.
This is where nonperturbative effects dominate.
This limits the extent of our theoretical prediction of the decay
rate to which the experimental data can be compared.
Therefore it is necessary to introduce a smearing in order to use
perturbation theory reliably~\cite{pqw}.
The size of the smearing is determined by requiring that the
nonperturbative effects be smaller than the parton-model result.

\section{Operator Product Expansion}
For processes involving $b$-quark decay,
it is appropriate to use the HQEFT.
In the HQEFT, the full QCD $b$ field is expressed in terms of
$b_v$ for $b$ quark velocity $v$.
The $b_v$ field is defined by
\begin{equation}
b_v = \frac{1+ v \hskip -.215cm / }{2} e^{im_b v \cdot x} b + \cdots,
\end{equation}
where the ellipses denote terms suppressed by powers of $1/m_b$.
The equation of motion for $b_v$ is $v \hskip-0.215cm / b_v = b_v$.
We can apply the techniques of the OPE to expand $T^{\mu \nu}$
in terms of matrix elements of local operators involving $b_v$ fields
\begin{eqnarray}
\displaystyle
T^{\mu \nu}
& = & -i \int d^4x e^{i q \cdot x}
\langle B | T\{J_u^{\mu}(x)^{\dagger} J_u^{\nu}(0)\} | B \rangle
\nonumber  \\
& = & \sum_{n,v} C_n^{\mu\nu} (v, q) \,
\langle B | {\cal O}_v^{(n)} | B \rangle.
\label{ope}
\end{eqnarray}
Here ${\cal O}_v^{(n)}$ are local operators involving
$\bar b_v b_v$ bilinears.
The coefficient functions $C^{\mu \nu}_n$ depend explicitly on
$q$ and $v$ because of the $v$ dependence of ${\cal O}_v^{(n)}$.
All large momenta are contained in the coefficient functions
while the matrix elements describe low-energy physics.

In this scheme $T^{\mu \nu}$ is expanded as a double series
in powers of $\alpha_s$ and $1 / m_b$.
To leading order in $\alpha_s$ and $1 / m_b$
\widetext
\begin{eqnarray}
T^{\mu\nu}  &=& -i \int \! d^4x \, e^{i (m_b v -q) \cdot x} \,
\langle B|T\{ {\overline b}_v(x) \gamma^{\mu}(1+\gamma_5 ) u(x)
{\overline u}(0) \gamma^{\nu} (1+\gamma_5)b_v (0) \} |B \rangle
\nonumber \\
 &=& -i \int \! d^4x \, e^{i (m_b v - q)  \cdot x} \,
 \langle B| T\{{\overline b}_v(x)
 \gamma^{\mu}(1+\gamma_5) S_0 (x) \gamma^{\nu} (1+\gamma_5)b_v(0)\}
 |B\rangle,
\label{tstart}
\end{eqnarray}
\narrowtext
\noindent where $S_0 (x)$ denotes a free $u$ quark propagator.
In momentum space Eq.~(\ref{tstart}) is proportional to
\begin{equation}
\gamma^{\mu} {{\cal Q} \hskip -.22cm / +k \hskip -.22cm /
\over ({\cal Q} +k)^2 } \gamma^\nu (1 + \gamma_5),
\label{uprop}
\end{equation}
where ${\cal Q} = m_b v - q $ is the momentum transferred to
the $u$ quark and $k^\mu$ is a small residual momentum of the
$b_v$ field of order $\Lambda$.

When $ \Lambda^2 \ll {\cal Q}^2 \ll m_b^2$,
it is sufficient to keep the terms at leading order in
$1 / m_b$ only and expand Eq.~(\ref{uprop}) in powers of
$k / |{\cal Q}|$. This generates coefficient functions and
local operators in which $k$ is replaced by the derivatives
acting on the $b_v$ fields.
The leading term independent of $k$ gives the parton-model result.
Contracting the leading term with $L_{\mu \nu}$ we find
\begin{eqnarray}
T^{\mu \nu}_0 L_{\mu \nu} & = &
64 {k_e \cdot {\cal Q} \over {\cal Q}^2} k_{\bar \nu}^{\mu}
\langle B | \bar b_v \gamma_\mu (1 + \gamma_5) b_v | B \rangle
\nonumber \\
&= & 128 m_b {k_e \cdot {\cal Q} k_{\bar \nu} \cdot v \over
{\cal Q}^2 },
\label{partonresult}
\end{eqnarray}
where $T^{\mu \nu}_0$ is $T^{\mu \nu}$ at leading order in
$k/|{\cal Q}|$.
The second line in Eq.~(\ref{partonresult})
follows from the normalization
$\langle B | \bar b_v b_v | B \rangle = 2 m_b$.

The next-order correction to  Eq.~(\ref{partonresult})
is suppressed by at least $(\Lambda / {\cal Q})^2$~\cite{chayetal}.
Corrections to $T^{\mu \nu}$ from higher dimensional operators
involving $\overline b_v b_v$ bilinears have been studied
systematically~\cite{bigietal,mannel,manoharwise}.
Furthermore Neubert~\cite{neubert} has resummed these
corrections to get a ``shape function''.
He has observed that the shape function
is universal independent of the final quark flavor.

Usually the coefficient functions are calculated perturbatively.
In addition there could be nonperturbative correction to the
coefficient functions as well~\cite{shifman}.
In the next section we calculate how instantons affect the
coefficient function of the leading term, Eq.~(\ref{partonresult}),
in the OPE.
The instantons contribute not only to the matrix elements but
also to the coefficient functions.
If there is an infrared divergence due to large instantons we
attribute it to the contribution to the matrix elements.
For infrared-finite parts we interpret them as the correction to
the coefficient functions.

\section{Instanton Contribution}
We now compute the contribution of instantons to the coefficient
functions.
In estimating the contribution we start from the Euclidean
region where ${\cal Q}^2$ is large enough to use the OPE reliably.
We expect that the main contribution comes from small instantons
of size $\rho
\mathrel{\rlap{\lower3pt\hbox{\hskip0pt$\sim$}}\raise1pt\hbox{$<$}}
{1 /|{\cal Q}|}$, hence we use the dilute gas approximation
in what follows.
More specifically we calculate the instanton correction to the
decay rate at leading order in both $\alpha_s$ and $k$ in the OPE.
This is the instanton correction to the parton model result
Eq.~(\ref{partonresult}).

In the background of an instanton ($+$ anti-instanton) of size $\rho$
and instanton orientation $U$ located at the origin, the Euclidean
fermion propagator may be expanded in small fermion masses
as~\cite{andreigross}
\begin{eqnarray}
\displaystyle
S_\pm (x, &y;& \rho_\pm; U_\pm)  =
- {1 \over m} \psi_0 (x) \psi^\dagger_0 (y)
+ S^{(1)}_\pm (x, y; \rho_\pm; U_\pm) \nonumber \\
&+& m \int d^4 w S^{(1)}_\pm (x, w; \rho_\pm; U_\pm)
S^{(1)}_\pm (w, y; \rho_\pm; U_\pm)
\nonumber \\
&+& {\cal O}(m^2),
\label{propexp}
\end{eqnarray}
where $\pm$ denotes instanton, anti-instanton.
$\psi_0$ is the fermion zero mode eigenfunction  and
$S^{(1)}_\pm = \sum_{E>0} {1 \over E} \Psi_{E \pm} (x)
\Psi^\dagger_{E \pm} (y)$ is the Green's function of fermion
nonzero modes.

In evaluating the forward Compton scattering amplitude
$T^{\mu \nu}$, Eq.~(\ref{tmunu}),
we should use the propagator in Eq.~(\ref{propexp})
instead of $S_0$. $T^{\mu \nu}$ can be written as
\begin{equation}
T^{\mu \nu}  = T^{\mu \nu}_0  + T^{\mu \nu}_{\rm inst.},
\label{tdecomp}
\end{equation}
where the first term is the parton-model amplitude.
The second term is the amplitude due to instantons of all
orientation $U$, position $z$ and size $\rho$.
A schematic configuration of an instanton or anti-instanton is
shown in Fig.2.
After averaging over instanton orientations $T^{\mu \nu}_{\rm inst}$
is given by
\widetext
\begin{equation}
T^{\mu \nu}_{\rm inst.}   =  \int \! d^4 \Delta \,
e^{i {\cal Q} \cdot \Delta } \sum_{a = \pm} \!\! \int
\! d^4 z_a \, d \rho_a D(\rho_a)
 \langle B | \bar b_v (x) \gamma^\mu (1+\gamma_5) \{ {\cal S}_a
(X, Y; \rho_a) -S_0 (\Delta)\} \gamma^\nu (1 + \gamma_5) b_v (y)
| B \rangle,
\label{insttmunu}
\end{equation}
\narrowtext
\noindent where $D(\rho)$ is the instanton density,
$\Delta = x - y, \,\,\, X = x - z$ and $Y = y - z$.
In Eq.~(\ref{insttmunu}), ${\cal S}_\pm (X, Y; \rho_\pm)$ is the
fermion propagator averaged over instanton (anti-instanton)
orientations centered at $z$.
Using the ${\overline {\rm MS}}$ scheme with $n_f$
flavors of light fermions, $D(\rho)$ is given by~\cite{bernard}
\begin{equation}
\displaystyle
D(\rho)  =  K \, \Lambda^5 \, (\rho \Lambda)^{6 + {n_f \over 3}}
      \, \biggl( \ln{1 \over \rho^2 \Lambda^2}
      \biggr)^{45-5 n_f \over 33 - 2 n_f},
      \label{density}
\end{equation}
where
\begin{eqnarray}
      \displaystyle
      K  & = & \biggl( \prod_i {\hat m_i \over \Lambda} \biggr) \,
2^{12 n_f \over 33 - 2n_f} \, \biggl({33 - 2n_f \over 12} \biggr)^6
      \nonumber \\
& \times & {2 \over \pi^2} \,
\exp\bigl[{1 \over 2} - \alpha (1) + 2 (n_f -1)
\alpha ({1 \over 2}) \bigr]
      \label{ddensity}
\end{eqnarray}
in which the $\beta$ function at two loops and the running mass
at one loop are used and $\hat m_i$ are the
renormalization-invariant quark masses.
In Eq.~(\ref{ddensity}) $\alpha(1) = 0.443307$ and
$\alpha(1/2) = 0.145873$.
{}From now on we replace the logarithmic
term in $D(\rho)$ by its value for $\rho = 1/|{\cal Q}|$.
Corrections to this replacement are
negligible since they are logarithmically suppressed.

Inserting Eq.~(\ref{propexp}) to Eq.~(\ref{insttmunu}),
the leading contribution comes from the mass-independent part
${\cal S}_\pm^{(1)}$ due to the chiral structure of the
left-handed weak currents.
The chirality-conserving part of ${\cal S}^{(1)}_\pm$
in singular gauge~\cite{andreigross} is written as
\widetext
\begin{eqnarray}
\displaystyle
\sum_{a = \pm} {\cal S}^{(1)}_a (X, Y; \rho) & = &
-{1 \over \pi^2} {\Delta \hskip-0.23cm / \over \Delta^{4}}
(X^2 Y^2 + \rho^2 X \cdot Y)[X^2 Y^2 (X^2 + \rho^2)
(Y^2 + \rho^2)]^{-{1 \over 2}} \nonumber \\
    && - {\rho^2  \over 4 \pi^2} {\Delta \hskip-0.23cm /
    \over \Delta^{2}}
    ( X^2 [X^2 Y^2 (X^2 + \rho^2)^3 (Y^2 + \rho^2)]^{-{1 \over 2}}
   + (X \leftrightarrow Y) ) \nonumber \\
    && + {\rho^2 \over 4 \pi^2} ( X \hskip-0.24cm / \,
    [X^2 Y^2 (X^2 + \rho^2)^3 (Y^2  + \rho^2)]^{-{1 \over 2}} -
    (X \leftrightarrow Y) ).
    \label{prop1}
    \end{eqnarray}
{}From Eq.~(\ref{prop1}) ${\cal S}_\pm(X, Y; \rho) - S_0(\Delta)$
can be written as
\begin{eqnarray}
\displaystyle
\sum_{a = \pm}
\bigl({\cal S}_a (X, Y; \rho)  - S_0 (\Delta) \bigr) &  \approx &
-{\rho^4 \over 8 \pi^2} {\Delta \hskip-0.23cm / \over \Delta^{4}}
   (X^2 - Y^2)^2 [X^2 Y^2 (X^2 + \rho^2)
   (Y^2 + \rho^2)]^{-1} \nonumber \\
   && + {\rho^4 \over 2 \pi^2} {\Delta \hskip-0.23cm / \over
   \Delta^{2}}
   \biggl({1 \over X^2 + \rho^2} + {1 \over Y^2 + \rho^2}\biggr)
 [X^2 Y^2 (X^2 + \rho^2)(Y^2 + \rho^2)]^{-{1 \over 2}} \nonumber \\
   &&
   + {\rho^2 \over 4 \pi^2} \biggl({{X \hskip -0.23cm /}
   \over X^2 + \rho^2} - {{Y \hskip -0.23cm /}
   \over Y^2 + \rho^2} \biggr)
   [X^2 Y^2 (X^2 + \rho^2 ) (Y^2 + \rho^2)]^{-{1 \over 2}},
\label{prop}
\end{eqnarray}
\narrowtext
\noindent
where $S_0 (\Delta) = -\Delta \hskip-0.23cm / / 2 \pi^2 \Delta^4$
is the free quark propagator.
In deriving Eq.~(\ref{prop}) we move $-\rho^2(X^2 + Y^2)/2$
proportional-part of the first line to the second line in
Eq.~(\ref{prop1}) using
$\rho^2 X \cdot Y = \rho^2 ((X+Y)^2 - X^2- Y^2)/2$.
We also approximate the remaining terms to get the first line in
Eq.~(\ref{prop})
so that exact analytic calculation is possible while leaving
small instanton contribution essentially unaltered.
Plugging Eq.~(\ref{prop}) into Eq.~(\ref{insttmunu}),
we integrate over the instanton center $z$,
the instanton size $\rho$, and finally make a Fourier
transform over $\Delta$.
Note that the integral over $z$ is convergent as
$|z| \rightarrow \infty$.

We notice that the integration over $\rho$ is convergent for small
instantons. On the other hand large instanton part is divergent.
However since the integrand is analytic for large $\rho$
(See Appendix.),
 there are only a finite number of divergent terms when
 $T^{\mu \nu}_{\rm inst}$ is expanded in $1 / \rho$.
 We interpret these infrared divergent terms as the instanton
 contribution to the matrix elements of operators in the
 OPE~\cite{balietal}. The remaining terms are infrared convergent
 and are interpreted that they contribute to the coefficient
 functions.
 In order to calculate finite
 terms any convenient regularization prescription will do.
To this end we analytically continue the exponent of $\rho$ in the
instanton density so that $D(\rho) \propto
\rho^{M-4}$~\cite{porrati} and the spacetime dimensions to
$d = 4 + 2 \epsilon$.
In the end we let $M \rightarrow 11$ for $n_f = 3$ and
$\epsilon \rightarrow 0$ to get the final answer.

After some algebra which we show in the Appendix,
the contribution to $T_{\rm inst}^{\mu \nu} L_{\mu \nu}$ from
the first line of Eq.~(\ref{prop}) consists of two pieces.
The contribution of the first piece is  written as
\begin{eqnarray}
&& K \! \int \! d \rho \, \rho^4 D(\rho) \int \!
{d^4 w \over w^2 (w^2 + \rho^2)} \nonumber \\
& \times &  2^6 \,
{k_e \cdot {\cal Q} \over {\cal Q}^2 } k^\mu_{\bar \nu}
\langle B | \overline b_v \gamma_\mu ( 1 + \gamma_5) b_v
| B \rangle.
\label{divterm1}
\end{eqnarray}
The integration over $\rho$ is  divergent as we take the
physical value $M = 11$.
This is interpreted as the large-instanton contribution to
the matrix element $\overline b_v \gamma^\mu b_v$.

The contribution from the second piece is regular as
$M \rightarrow 11$, $\epsilon \rightarrow 0$ and is given by
\begin{eqnarray}
\displaystyle
 && K \, {2^{10} \over 35}\, \pi^2 \Gamma^2 (6) \,
\, \biggl({\Lambda^2 \over {\cal Q}^2} \biggr)^6 \,
\biggl( \ln {{\cal Q}^2 \over \Lambda^2} \biggr)^{10 / 9}
\nonumber \\
& \times & 2^6 \,
{k_e \cdot {\cal Q} \over {\cal Q}^2 } k^\mu_{\bar \nu}
\langle B | \overline b_v \gamma_\mu ( 1 + \gamma_5) b_v
| B \rangle.
\label{convterm1}
\end{eqnarray}
The contribution from the second and the third lines
in~Eq.(\ref{prop})
to $T^{\mu \nu}_{\rm inst} L_{\mu \nu}$  vanishes,
as explained in the Appendix.
Therefore Eq.~(\ref{convterm1}) is the overall finite contribution.

It is now straightforward to evaluate the instanton contribution
to the differential decay rate.
So far we have worked in Euclidean spacetime.
We now make a naive analytic continuation to Minkowski spacetime
with timelike ${\cal Q}^2 = (m_b v - q)^2$ to evaluate the
differential decay rate.
Then $T^{\mu \nu}_{\rm inst}L_{\mu \nu}$ is written as
\begin{eqnarray}
\displaystyle
T^{\mu \nu}_{\rm inst} L_{\mu \nu} & = & - {5 \over 4} A m_b^2 \,
{(v \cdot \hat q - {y \over 2}) (y - \hat q^2) \over
(v \cdot \hat q - {1 + \hat q^2 \over 2} )^7} \nonumber \\
 & \times & \biggl[\ln {-2 m_b^2 \over \Lambda^2}
\bigl( v \cdot \hat q - {1 + \hat q^2 \over 2} \bigr) \biggr]^{10/9},
\label{timunu}
\end{eqnarray}
where
\begin{equation}
A = {2^{11} \pi^2 \over 175} \, \Gamma^2(6) \, K
\biggl({\Lambda \over m_b} \biggr)^{12}.
\label{a}
\end{equation}
Eq.~(\ref{timunu}) has a branch cut emanating from the
resonance point $v \cdot \hat q = {1 + \hat q^2} / 2$,
which we have shown in Fig.1.

To get the double differential decay rate
$d^2 \Gamma / d\hat q^2 d y$,
we integrate  Eq.~(\ref{timunu}) over
$v \cdot \hat q$ along $C$ in Fig.1 as described in Section 2.
The differential decay rate is written as
\begin{eqnarray}
\displaystyle
{1 \over \Gamma_0} {d \Gamma_{\rm inst} \over  d \hat q^2 d y}
& = & {15 \over 32 \pi} A \,
\int_C  d v \cdot \hat q \, {(y - \hat q^2)
(v \cdot \hat q - {y \over 2})
\over ( v \cdot \hat q - {1 + \hat q^2 \over 2})^7} \nonumber \\
& \times & \biggl[\ln {-2 m_b^2 \over \Lambda^2}
\bigl( v \cdot \hat q - {1 + \hat q^2 \over 2} \bigr)
\biggr]^{10/9},
\label{ddrate}
\end{eqnarray}
where $\Gamma_0 = G_F^2 |V_{ub}|^2 m_b^5 / 192 \pi^3$.
The exponent of the logarithm is almost unity and we replace $
10/9$ by $1$. We expect that it does not change the result much.
We finally get
\begin{equation}
\displaystyle
{1 \over \Gamma_0} {d^2 \Gamma_{\rm inst} \over d \hat q^2 d y} =
A \, y^5 \,
{ 5 \hat q^2 - (1-y) (y-\hat q^2) \over (1 - y)^{6}
(y - \hat q^2)^{5}}.
\label{ddrate2}
\end{equation}
The instanton effect is suppressed at $y \sim 0$.
However it become singular near the boundaries of the phase space
at $y \sim 1$ and $y \sim \hat q^2$.
At these boundaries the final quark approaches the resonance region.
As there are large instanton contributions near the resonance point,
it is necessary to introduce a smearing to define sensible
decay rate in perturbative QCD.
We discuss this in detail in the next section.

\section{Numerical Analysis}
We have calculated the instanton effect on the inclusive
semileptonic $b$ decay.
As we have observed in the previous section the effect is
singular at the boundaries of the phase space.
In order to use perturbative expansions
reliably a smearing prescription has to be introduced.

As a simple prescription we first consider the smeared single
differential decay rate
\begin{equation}
\langle {d \Gamma_{\rm inst} \over d y} \rangle_\delta
= \int_0^y d \hat q^2 \, \theta( y - \hat q^2 - \delta)
{d \Gamma_{\rm inst} \over d y}
\label{singsmear}
\end{equation}
in which we restrict the phase space so that the singular region
$y = \hat q^2$ is avoided by $\delta$.
The size of smearing $\delta$ is determined by the requirement
that the smeared instanton contribution to
the single differential decay rate be smaller than that in
the parton model
$d \Gamma_0 / d y = 2 y^2 (3 - 2 y) \Gamma_0$.
Eq.(\ref{singsmear}) is given by
\begin{eqnarray}
\langle {d \Gamma_{\rm inst} \over d y} \rangle_\delta
& = & \bigl({d \Gamma_0 \over d y} \big)
{ A \over 24} {1 \over (1-y)^6 (3 - 2y)}
\nonumber \\
& \times & \bigl[9 - 4 y - 24({y \over \delta})^3 + 15
({y \over \delta})^4 + 4 ({y^4 \over \delta^3}) \bigr].
\label{singsmear2}
\end{eqnarray}

The ratio $R(y, \delta) = \langle d \Gamma_{\rm inst}
/ dy \rangle_\delta / (d \Gamma_0 / dy)$
is plotted in Fig.3 for three different values of
$\delta$ = 0.15, 0.17, 0.19 with $\Lambda = 400$ MeV and
$m_b = 5$ GeV.
The numerical value of $A$ is given by
\begin{equation}
A = \biggl({19.2 {\rm GeV} \over m_b} \biggr)^3 \,
\biggl({\Lambda \over m_b} \biggr)^9,
\label{numa}
\end{equation}
in which we set $n_f = 3$ and the renormalization-invariant
quark masses as~\cite{quarkmass}
 \begin{eqnarray}
 \hat m_u & =& 8.2 \pm 1.5 {\rm MeV}, \nonumber \\
 \hat m_d & = & 14.4 \pm 1.5 {\rm MeV}, \nonumber \\
 \hat m_s & = & 288 \pm 48 {\rm MeV}.
 \end{eqnarray}
We see that for these choices of $\delta$, $R (y, \delta)$
grows rapidly for $y
\mathrel{\rlap{\lower4pt\hbox{\hskip1pt$\sim$}}\raise1pt\hbox{$>$}}
0.84$. From Fig.3 we see that a smearing of
$ \delta \approx 0.16 \sim 0.20$ at the boundary
$y = \hat q^2$ of the phas space is needed to extract a
reliable electron spectrum from the single differential decay rate.

We have also examined the behavior of $R(y, \delta)$ as we increase
$\delta$.
The position at which $R(y, \delta)$ grows like a brick wall
does not shift much from $y \sim 0.84$. Because of this singular
behavior we need another smearing near $y = 1$.
For simplicity we cut off the region near $y = 1$ by the
same smearing width $\delta$ as in Eq.~(\ref{singsmear}).

Instanton contribution to the total decay rate after such a smearing
prescription is given by
\begin{eqnarray}
\displaystyle
\langle  \Gamma_{\rm inst} \rangle_\delta & =&
\! \int_0^1 \! d y \! \int_0^y \! d \hat q^2
\theta(y-\hat q^2 -\delta) \theta(1 - y - \delta)
{d^2 \Gamma_{\rm inst} \over d \hat q^2 d y} \nonumber \\
& = &  \Gamma_0 \, {A \over 24}
\, {1 \over \delta^9}
( \, 6 - 53 \delta + 198 \delta^2 - 420 \delta^3
+ 612 \delta^4  \nonumber \\
&& - 354 \delta^5 + 16 \delta^6 - 4 \delta^7 - \delta^9
+ 180 \delta^5 \ln \delta).
\label{totsmear}
\end{eqnarray}
The ratio $R(\delta) = \langle \Gamma \rangle_\delta / \Gamma_0$
is plotted in Fig.4 for three different values of $\Lambda =
350, 400, 450$ MeV and for $m_b = 5$ GeV.
We see that the ratio $R (\delta)$ also rises sharply like
a brick wall as $\delta$ decreases.
When $\delta$ is large, say $\delta
\mathrel{\rlap{\lower4pt \hbox{\hskip1pt$\sim$}}\raise1pt\hbox{$>$}}
0.15$, $R (\delta)$ is
insensitive to the choice of $\Lambda / m_b$. However, for small
$\delta$, say $ \delta \approx 0.12$, $R(\delta)$ is sensitive to
the value of $\Lambda / m_b$. If we require $R (\delta )$ be less
than 20\%,
the size of the smearing is roughly at least $ 0.12 \sim 0.15$
for our choices of $\Lambda$.

Our results Eqs.~(\ref{singsmear2}),~(\ref{totsmear}) are
sensitive to the precise value of $\Lambda$.
This is what we expect from the instanton effect.
Nevertheless, as emphasized in Introduction, our results may
still represent a typical size of nonperturbative effects.

\section{Conclusion}
In this paper we have studied instanton effects on the decay rate of
the inclusive semileptonic $b \rightarrow u$ decay.
In particular we have estimated the instanton correction to the
coefficient function at leading order.
We have found that the correction is singular at the boundaries
of the phase space.
Because of the large instanton contribution to the
coefficient function,
we have found that it is necessary to introduce a smearing near the
boundaries of the phase space.
For the single differential decay rate a smearing of size
$\delta \approx 0.16 \sim 0.2$ is needed taking into account
of the correlation of the
smearing sizes at the two boundaries $ y = 1$ and $ y = \hat q^2$.
For the total decay rate the smearing size is
$\delta \approx 0.12 \sim 0.15$. Note that $d \Gamma_0 / d y$ is
appreciable for all $y$ while $d \Gamma_{\rm inst} / d y$ is
negligible except at the boundaries of the Dalitz plot.
Therefore the smearing size for the total decay rate is smaller
than that for the single differential decay rate.

Precise theoretical prediction in the endpoint region is important
for the extraction of $V_{ub}$.
Singular nature of various corrections indicates we need some
smearing before comparing with data.
Manohar and Wise~\cite{manoharwise} have studied the smearing
necessary for the effects of matrix elements of higher
dimensional operators.
They have concluded that the size of smearing should be
$\epsilon \sim 0.2$.
Neubert ~\cite{neubert} has suggested an idea to measure
the universal shape function from $b \rightarrow c$ decay and
apply it to $b \rightarrow u$ decay.
He has proposed this as a promising way to a precise extraction
of $V_{ub}$.

Falk et.al.~\cite{falketal} have studied the smearing for the
validity of the leading log approximation and have found
the smearing size should be roughly $0.1$ or less.
The instanton effect is another source of corrections to $B$ decays.
While it is negligible for $b \rightarrow c$ decay the
effect is significant in $b \rightarrow u$ decay.
Our analysis shows that the smearing has to be about
$\delta \approx 0.12 \sim 0.16$.
This is comparable to the difference in the endpoints of the
$b \rightarrow c$ and $b \rightarrow u$ decays.
Therefore it is perhaps difficult to eliminate a model
dependence in extracting $V_{ub}$ from inclusive semileptonic
$b \rightarrow u$ decay.

We can similarly consider the instanton effect in inclusive
semileptonic $c$ decay.
As we can infer from the quark mass dependence in Eq.~(\ref{a})
we expect the effect is much larger in this case because
$m_c \ll m_b$.
Therefore the decay rate near the endpoint is strongly
model-dependent due to large uncertainties from radiative
corrections and nonperturbative effects~\cite{manoharwise}
including instantons.
We may also consider the instanton effect on inclusive
$b \rightarrow c$ decay.
Because ${\cal Q}^2
\mathrel{\rlap{\lower4pt\hbox{\hskip1pt$\sim$}}\raise1pt\hbox{$>$}}
m_c^2$ the instanton effect is finite even at the boundaries
of the phase space.
The numerical value is negligibly small since it is proportional to
high powers $(\Lambda / m_c)^n$ where $n$ may be determined from the
instanton density Eq.~(\ref{density}) for $n_f = 4$.

\section*{Acknowledgements}
SJR thanks M. Luke, M. Peskin, J. Preskill, H. Quinn, M. Shifman and
M.B. Wise for useful discussions.
This work was supported in part by KOSEF-SRC program (SJR, JGC),
KOSEF Grant`94 (JGC), Ministry of Education through SNU-RIBS (SJR)
and BSRI-93-218 (JGC), KRF-Nondirected Research Grant`93 (SJR).

\section*{Appendix: Instanton Integrations}
There are three integrations [See Eq.~(\ref{prop}).]
we have to perform.
The integrals of Eq.~(\ref{prop}) over $\rho, z$
and Fourier transform of $\Delta$ after a suitable change of
variables are expressed as
\begin{eqnarray}
\displaystyle
&& K \biggl(\ln {{\cal Q}^2 \over \Lambda^2} \biggr)^{10/9}
\int_0^\infty d \rho \, \rho^{M-4} \int d^d t \, d^d u {\rho^2
\over 4 \pi^2}
\nonumber \\
&\times& \biggl[ \biggl( {i \partial \over {\partial {\cal Q}
   \hskip-0.23cm / }} {e^{i {\cal Q} \cdot t} \over t^4} \biggr)
  \biggl(
  {\rho^2 \over u^2 (u^2 + \rho^2)} -
  { \rho^2 \over u^2 ((t-u)^2 + \rho^2)}
  \biggr)  \label{first}  \\
&-&
\biggl( {i \partial \over {\partial {\cal Q} \hskip-0.23cm / }}
{e^{i {\cal Q} \cdot t} \over t^2} \biggr) \,
{2 \rho^2 \over \sqrt{u^2 (u^2 + \rho^2)
(t-u)^2 ((t-u)^2 + \rho^2)^3}}
\label{second} \\
&-& \,\,
\biggl(2 {i \partial \over {\partial {\cal Q} \hskip-0.23cm / }}
{e^{i {\cal Q} \cdot t} \over \sqrt{t^2 (t^2 + \rho^2)^3}}
\biggr) \,\,
 {e^{i{\cal Q}\cdot u} \over \sqrt{u^2 (u^2 + \rho^2)}} \,\, \biggr],
\label{third}
\end{eqnarray}
where ${\partial / \partial {\cal Q} \hskip-0.23cm / }
= \gamma_\mu {\partial / \partial {\cal Q}^\mu}$.
At small $\rho$, the $\rho$ integral is convergent while  at
large $\rho$ it is divergent.
However for large $\rho$ the integrand may be expanded in powers of
$1 / \rho$.
We can see clearly that there are only a finite number of
divergent terms at the physical value $M=11$ for $n_f=3$.
These divergent terms are interpreted as the instanton contribution
to the matrix elements of operators in the OPE~\cite{balietal}.
Since the remaining terms are finite we may evaluate them by any
convenient methods. We have analytically continued $M$ and
the spacetime dimensions $d = 4 + 2 \epsilon$.
We set $M = 11$ and $d=4$ in the end.

The first term in Eq.~(\ref{first}) is divergent and we interpret
it as a contribution to the hadronic matrix
elements~\cite{balietal}.
In evaluating the second term in Eq.~(\ref{first})
we first integrate over $u$ and then over $\rho$.
Finally we make a Fourier transform with respect to $t$.
The result turns out to be regular as $\epsilon \rightarrow 0$.
Setting $\epsilon =0$ the second term in  Eq.~(\ref{first})
is given by
\begin{eqnarray}
&& K \, 2^{M-1}\pi^2 \frac{M+1}{M+3}
\Gamma^2 \biggl(\frac{M+1}{2} \biggr)
\frac{\Gamma(-\frac{1+M}{2})}{\Gamma(-\frac{M-3}{2})} \nonumber \\
&& \times \frac{i{\cal Q}
\hskip -0.24cm /}{{\cal Q}^{M+3}}\,
\biggl( \ln {{\cal Q}^2 \over \Lambda^2} \biggr)^{10/9}.
\label{term1int}
\end{eqnarray}
In getting Eq.~(\ref{term1int}) we have analytically continued
$M$ in such a way that the integral becomes convergent.
However the result can be extended to $M=11$ as it is
convergent for any positive $M \ge 3$.
The final result is
\begin{equation}
K \, \frac{2^{M+1}\pi^2}{(M+3)(M-1)}
\Gamma^2 \biggl(\frac{M+1}{2} \biggr)
\frac{i{\cal Q}\hskip -0.22cm /}{{\cal Q}^{M+3}}
\,  \biggl(\ln {{\cal Q}^2 \over \Lambda^2} \biggr)^{10/9}.
\end{equation}
Putting $M=11$ yields
\begin{equation}
 4.16\times 10^6 \,K\, \frac{i{\cal Q}\hskip -0.22cm /}
 {{\cal Q}^{14}}
 \,  \biggl( \ln {{\cal Q}^2 \over \Lambda^2} \biggr)^{10/9}.
\label{result1}
\end{equation}

Eq.~(\ref{second}) is regular as $\epsilon \rightarrow 0$.
Setting $\epsilon = 0$ and after some calculation it is written as
\begin{eqnarray}
\displaystyle
&& K \,  2^{M-2}\pi^2 \frac{1+M}{1-M}\Gamma^2
\biggl(\frac{M+1}{2} \biggr)
\frac{i{\cal Q}\hskip -0.24cm /}{{\cal Q}^{M+3}}
\, \biggl( \ln {{\cal Q}^2 \over \Lambda^2} \biggr)^{10/9}
\nonumber \\
&\times& \! \int \!\! \int_0^1 \! dx \,dy \,
x^{-(M+2)/2} (1-x)^{-1/2} y^{1/2} (1-y)^{-1/2}
\nonumber \\
& \times & F \bigl(\frac{M+1}{2}, \frac{M+3}{2};M+2;
1-\frac{y}{x} \bigr),
\label{term2int}
\end{eqnarray}
where $F \equiv \, _2F_1$ is the hypergeometric function.
Here $x$ and $y$ are Feynman parameters introduced in
integrating over the instanton position.
Eq.~(\ref{term2int}) is also regular as
$M\rightarrow 11$, hence we can set $M=11$.
We can expand the hypergeometric function as a series of $(y/x)^n$
for $n\geq 0$. The $y$ integral is of the form
\begin{equation}
\int_0^1 dy y^{n+1/2}(1-y)^{-1/2},
\end{equation}
which is finite for all $n\geq 0$.
On the other hand the $x$ integral is of the form
\begin{equation}
\int_0^1 dx x^{-M/2-n-1}(1-x)^{-1/2},
\label{xint}
\end{equation}
which is zero for odd $M$ and infinity for even $M$.
Therefore for $M=11$ the integral vanishes.

Similarly Eq.~(\ref{third}) is regular as
$\epsilon\rightarrow 0$ and $M \rightarrow 11$.
After some calculation the result with $\epsilon=0$, $M=11$
is written as
\begin{eqnarray}
\displaystyle
&& K \, 2^{12} { \Gamma(7) \Gamma^2(6) \Gamma(5) \over \Gamma(12)}
\, \frac{i{\cal Q}\hskip -0.24cm /}{{\cal Q}^{14}}
 \, \biggl( \ln {{\cal Q}^2 \over \Lambda^2} \biggr)^{10/9}
 \nonumber \\
&\times& \int_0^1 dx dy x^{-11/2}(1-x)^{-1/2}y^{-1/2}(1-y)^{1/2}
\nonumber \\
& \times & F \bigl(5,6;12;1-\frac{y}{x} \bigr).
\end{eqnarray}
Again the hypergeometric function can be expanded as a
series of $(y/x)^n$ for $n \ge 0$.
The $y$ integral is finite but the $x$ integral vanishes for
the same reason as in Eq.~(\ref{xint}).
Therefore it turns out that Eqs.~(\ref{second}), (\ref{third})
vanish and the only finite contribution is Eq.~(\ref{result1}).

\begin{figure}
\caption{Anaytic structure of $T^{\mu \nu}$ in the complex
$ v \cdot \hat q$ plane.
The point $P$ is the minimum of $v \cdot \hat q$ at
$y/2 + \hat q^2 / 2y$ for $b \rightarrow u$ decay.
The contour integral along $C^{\prime}$ is related to the
contour integral along $C$.}
\label{fig1}
\end{figure}

\begin{figure}
\caption{Schematic diagram of an instanton configuration with
size $\rho$ located at $z_\mu$ contributing to the forward
Compton amplitude.}
\label{fig2}
\end{figure}

\begin{figure}
\caption{$R(y, \delta)$ for $\delta$= 0.15, 0.17, 0.19 with
$\Lambda=400$ MeV and $m_b$=5 GeV.}
\label{fig3}
\end{figure}

\begin{figure}
\caption{$R(\delta)$ for $\Lambda$= 350, 400, 450 MeV with
$m_b$= 5 GeV.}
\label{fig4}
\end{figure}

\end{document}